# Integration of ontology with machine learning to predict the presence of covid-19 based on symptoms

**Hakim El Massari[1], Noreddine Gherabi[1], Sajida Mhammedi[1], Hamza Ghandi[1], Fatima Qanouni[1], Mohamed Bahaj[2]**
[1]National School of Applied Sciences, Sultan Moulay Slimane University, Lasti Laboratory, Khouribga, Morocco
[2]Faculty of Sciences and Technologies, Hassan First University, Settat, Morocco

**ABSTRACT**

Coronavirus (covid 19) is one of the most dangerous viruses that have spread all over the world. With the increasing number of cases infected with the coronavirus, it has become necessary to address this epidemic by all available means. Detection of the covid-19 is currently one of the world's most difficult challenges. Data science and machine learning (ML), for example, can aid in the battle against this pandemic. Furthermore, various research published in this direction proves that ML techniques can identify illness and viral infections more precisely, allowing patients' diseases to be detected at an earlier stage. In this paper, we will present how ontologies can aid in predicting the presence of covid-19 based on symptoms. The integration of ontology and ML is achieved by implementing rules of the decision tree algorithm into ontology reasoner. In addition, we compared the outcomes with various ML classifications used to make predictions. The findings are assessed using performance measures generated from the confusion matrix, such as F-measure, accuracy, precision, and recall. The ontology surpassed all ML algorithms with high accuracy value of 97.4%, according to the results.

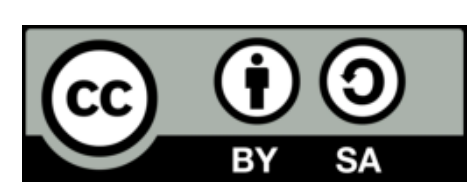

***Corresponding Author:***

Hakim El Massari
National School of Applied Sciences, Sultan Moulay Slimane University, Lasti Laboratory
Khouribga, Morocco
Email: h.elmassari@usms.ma

## 1. INTRODUCTION

Coronavirus (covid 19) emerged in 2019, called severe acute respiratory syndrome coronavirus 2 (SARS-COV-2). The World Health Organization (WHO) declared the covid 19 to be a worldwide pandemic in March 2020, and the data confirmed that covid 19 is transmitted from one person to another by mixing and proximity between people at a distance of about two meters. As for the method of spread, it is by respiratory droplets, when a person infected with the virus sneezes, coughs or breathes and another person close to him inhales it or enters his mouth, nose or eyes, and the coronavirus can also be transmitted by a person infected but not showing symptoms, or by airborne, or by touching a surface covered with the virus and then directly touching the person's mouth, nose or eyes. Symptoms of the covid-19 virus appear within 14 days of exposure to the virus and include fever, cough, loss of smell or taste, shortness of breath, muscle aches, nausea, diarrhea, sore throat, chest pain and chills. The U.S. Food and Drug Administration relied on the use of the vaccine to prevent infection with the covid 19 virus, so it relied on the use of Pfizer-Biontech antiviral for people aged 15 years and older, and its use in emergency situations for children aged 5 to 15 years, and for people aged 18 years More than that, the U.S. Food and Drug Administration relied on the use of the Moderna vaccine to prevent them from infection with the covid 19 virus. As for people with chronic illnesses, they should consult a doctor about other ways to protect themselves from covid 19 infection. To avoid covid 19 infection, the WHO and the center for disease control and prevention have recommended that certain precautionary measures be taken, including the following:

- Avoid gatherings in crowded, closed areas and keep a distance of two meters between you and others.
- Receive the vaccine.
- Washing hands with soap and water for 20 seconds, or using an alcohol-based hand sanitizer with at least a 60% concentration.
- Wear a mask in enclosed public places.
- Clean and disinfect surfaces.
- When coughing or sneezing, avoid touching your eyes, nose, or mouth and cover the nose and mouth with a tissue or elbow.

Machine learning (ML) is one of the most constantly evolving areas of com-puter science, with a wide range of applications. It is the process of obtaining usable information from a big quantity of data [1]. Marketing, industry, medical diagnosis, and other scientific domains all make use of ML approaches. ML algorithms are well-suited for medical data analysis since they have been frequently employed in medical datasets. ML comes in several forms, including classification, regression, and clustering. Each form has a particular consequence and influence depending on the problem that we are attempting to address. We focus on classification algorithms in our work because of their high accuracy and performance in classifying a given dataset into predetermined categories and predicting future events or information from that data. In the medical field, classification algorithms are often utilized, particularly in the diagnosis of illnesses such as covid-19. As a result, regularly used machine learning classification methods such as support vector machine (SVM), k-nearest neighbours (KNN), artificial neural network (ANN), Naive Bayes (NB), logistic regression (LR), and decision tree (DT) are utilized to detect patients with coronavirus at an early stage.

Recently, researchers published a significant quantity of research utilizing machine-learning algorithms to diagnose covid-19 [2]–[4]. In this comparative analysis [5], authors aim to determine which Classification technique has the highest accuracy rate for covid-19 positive data samples collected, the outcomes give 85%, 80%, and 65% of accuracy, for SVM, KNN, and NB respectively. Another study's [6] authors conducted an analysis based on incidents that occurred in different states of India in chronological order. They performed data cleansing and feature selection on the dataset, followed by forecasting of all classes using neural network, SVM, linear model, random forest (RF), and DT, where the RF model outperformed the others. SVM algorithm is the more efficient algorithm in predicting covid-19, with an accuracy of 98.81% [7].

Several researches have been conducted, and various machine learning models have been implemented, to identify and predict coronavirus diagnoses [8], [9]. Furthermore, many research to predict covid-19 illnesses have been completed, and numerous machine learning models have been implemented [10], [11] with the objective of categorizing and predicting coronavirus disease diagnoses. Various methods, like RF algorithm, SVM, genetic algorithm (GA), and ANN, particle swarm optimization (PSO), DT, NB and KNN were applied on a covid-19 dataset to predict the presence of coronavirus disease [12], [13]. Following that, multiple ML models were trained to predict covid-19. Furthermore, the metrics resulting from the confusion matrix were produced to evaluate the models' performance.

Covid-19 diagnosis and prediction has garnered a lot of attention in recent years, and numerous ways have been taken to address this issue [14]–[17]. To forecast the total number of confirmed cases the authors [18] proposed a model composed of SVM regression, polynomial regression, and linear regression. [19] At an early stage, machine learning algorithms are used on a dataset of Mexican patients to evaluate the severity of the condition based on their chronic disorders. This research combines several machine learning techniques such as DT, RF, BS, KNN, and LR to determine the influence of lifelong diseases on enhancing the symptoms of the virus. In this study [20], authors used the prophet method to forecast Covid-19 spread over the following year.

Furthermore, ontology has been one of the most widely used techniques to managing, organizing, and extracting data throughout the last few decades. It is a way of data representation that has been effectively utilized in a number of domains, particularly the medical domain. It is significant in computer science because of its ability to express many concepts and their relationships across fields. In reality, no single ontology is sufficient to meet today's expanding healthcare demands, and ontologies must be combined with machine learning algorithms to facilitate data integration and analysis. Massari *et al*. [21] created and explored an ontology-based DT model able to predict diabetes, [22] then compared the findings to numerous ML techniques, and discovered that the ontology model outperforms all other classifiers.

In this research, we intend to compare seven prominent classification approaches with the ontological model using carefully chosen criteria obtained from the confusion matrix, including F-measure, accuracy, recall, and precision. The rest of this paper is organized as follows: section 2 describes the methodologies utilized in this comparison analysis. Section 3 summarizes the findings and discussion. Section 4 concludes and discusses future work.

## 2. METHODS AND EVALUATION

The approaches and materials employed, as well as the experimental methodology, dataset description, machine learning algorithms, ontology model, and evaluation metrics, are all included in this section. Figure 1 depicts the process flowchart for this comparative study.

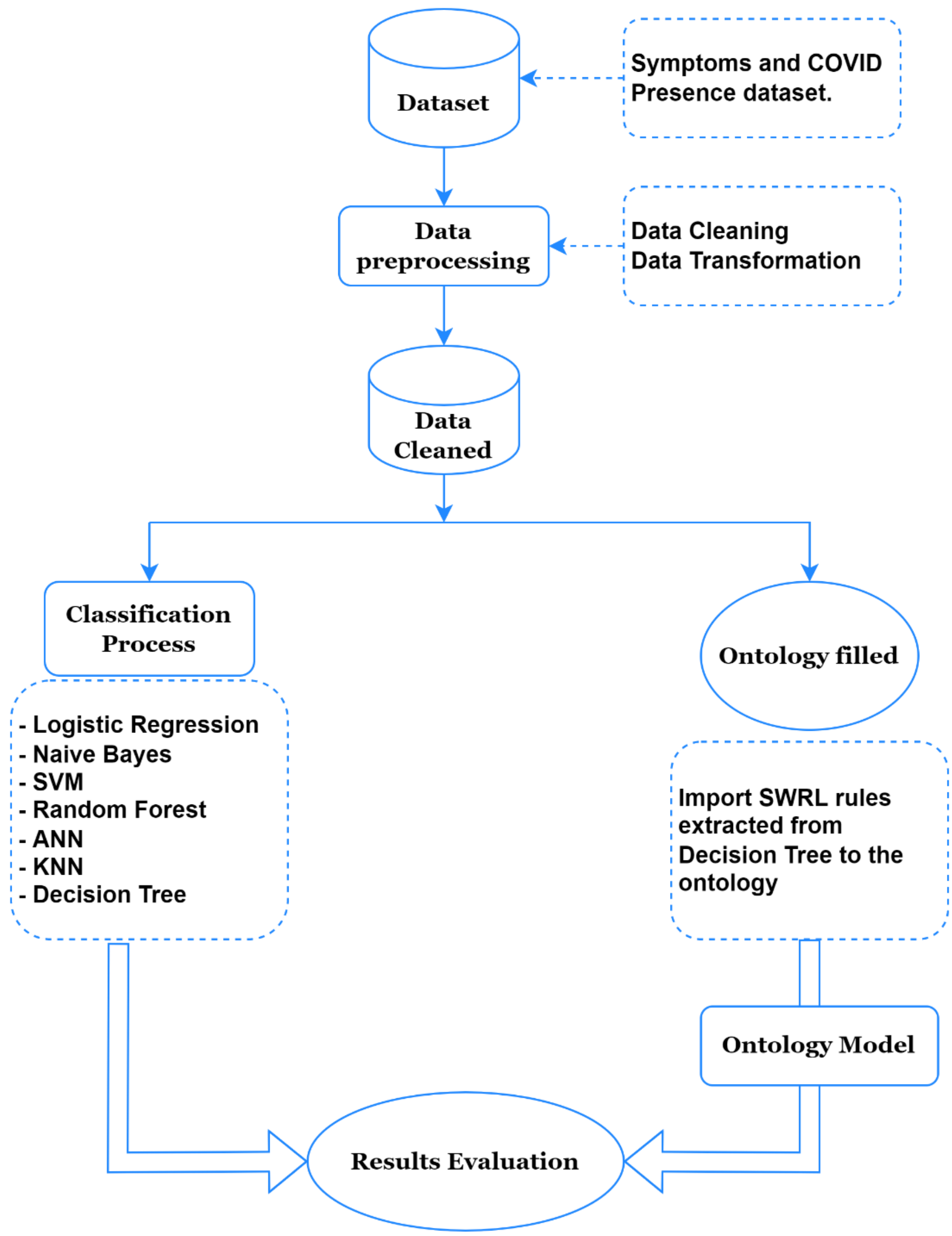

Figure 1. Experimental workflow

### 2.1. Data preprocessing

The dataset used is Symptoms and covid-19 Presence from Kaggle website [23], it consists of 5434 instances and 21 features (20 attributes and the last one is a target). A full description of all dataset attributes is provided in Table 1.

Table 1. Dataset feature's information

| Attribute | Description |
|---|---|
| 1- breath_pro | Breathing problem: the individual is having trouble breathing. |
| 2- fever | The temperature is higher than typical. |
| 3- dry_cough | Coughing that isn't accompanied by phlegm. |
| 4- sore_throat | The individual is suffering from a sore throat. |
| 5- run_nose | Running nose: the individual is suffering from a runny nose. |
| 6- asthma | The individual suffers from asthma. |
| 7- cld | Chronic lung disease: the individual suffers from lung illness. |
| 8- headache | The individual is suffering from a headache. |
| 9- heart_disease | The individual suffers from cardiovascular illness. |
| 10- diabetes | The individual has diabetes or has a family history of diabetes. |
| 11- hyper_tension | having an elevated blood pressure. |
| 12- fatigue | The individual is fatigued. |
| 13- gastrointestinal | The individual having some gastric issues. |
| 14- abroad_travel | Recently traveled outside the country. |
| 15- ccp | Contact with covid patient: interaction with a covid-19 infected individual. |
| 16- alg | Attended large gathering. |
| 17- vpep | Visited public exposed places. |
| 18- fwpep | Family working in public exposed places. |
| 19- wearing_masks | Wearing masks. |
| 20- sm | Sanitization from market: before using things purchased from the market, they should be sanitized. |
| 21- covid-19 | Predicted class (presence or absence of coronavirus). |

To build an effective machine learning classifier, we should always start with data cleaning, normalization of features, transformation of features, and even creation of new features from the dataset. The dataset contains 4968 similar instances, after removing duplicated instances the remaining is 466 instances, where 385 represents individuals with covid-19 and 81 represents individuals without covid-19. We would like to inform you that in order to provide a fair comparison of the classification results obtained, we did not use any feature selection or performance-boosting methods.

### 2.2. Machine learning algorithms

We have used weka software for all machine learning algorithms to predict whether the individuals have coronavirus or not. Weka comprises tools for data classification, clustering, visualization, preparation, association rules mining, and regression.

We used the seven most classifiers used to classify datasets (DT, RF, LR, ANN, NB, SVM, KNN. In addition, we employed two modes of test options: 10-fold crossvalidation and percentage split (split 70% train, remainder test) for the reason of enriching the study.

### 2.3. Ontological model

This section presents the technologies used to create the ontology, besides the approach used to build the ontology model with the help of rules extracted from DT. This methodology has been referred to in this research for more details [24], which we recommend reading for more information. We'll go through some specifics shortly here.

#### 2.3.1. Ontology construction

The ontology was built using the Protégé software, which is an open-source platform that provides a set of tools to a growing user community for constructing domain models and knowledge-based applications with ontologies [25]. The ontology was created manually; the main classes are diagnostic and patient. The graphical representation of the ontology is shown in Figure 2.

#### 2.3.2. Data properties and instances

The data properties used in the ontology are the same attributes presented in Table 1 which are used to build models of machine learning algorithms. Figure 3 illustrates the data properties. A plugin among the Protégé software plugins called Cellfie is used to import the same dataset used in Weka.

#### 2.3.3. Semantic web language rules and pellet reasoner

Following the creation of classes, data properties, and instances in the ontology. We need to establish the semantic web language rules (SWRL) reasoning rules. To achieve this, we used the SWRLTab plugin, we retrieved created rules from the DT algorithm and imported them into Protégé. The collected rules from the DT algorithm are converted using the Java programming language, with each leaf of the tree extracted as a single SWRL rule. For instance:

A leaf from the DT algorithm

*If breath_pro = No && sore_throat = Yes && dry_cough = Yes && fever = Yes THEN put the individual in presence*

SWRL resulted

*Patient(?pt) ^ breath_pro(?pt, ?Br) ^ swrlb:equal(?Br, 'No'^^xsd:string) ^ sore_throat(?pt, ?ST) ^ swrlb:equal(?ST, 'Yes'^^xsd: string) ^ dry_cough(?pt, ?DC) ^ swrlb:equal(?DC, 'Yes'^^xsd: string) ^ fever(?pt, ?F) ^ swrlb:equal(?F, 'Yes'^^xsd: string) → presence*

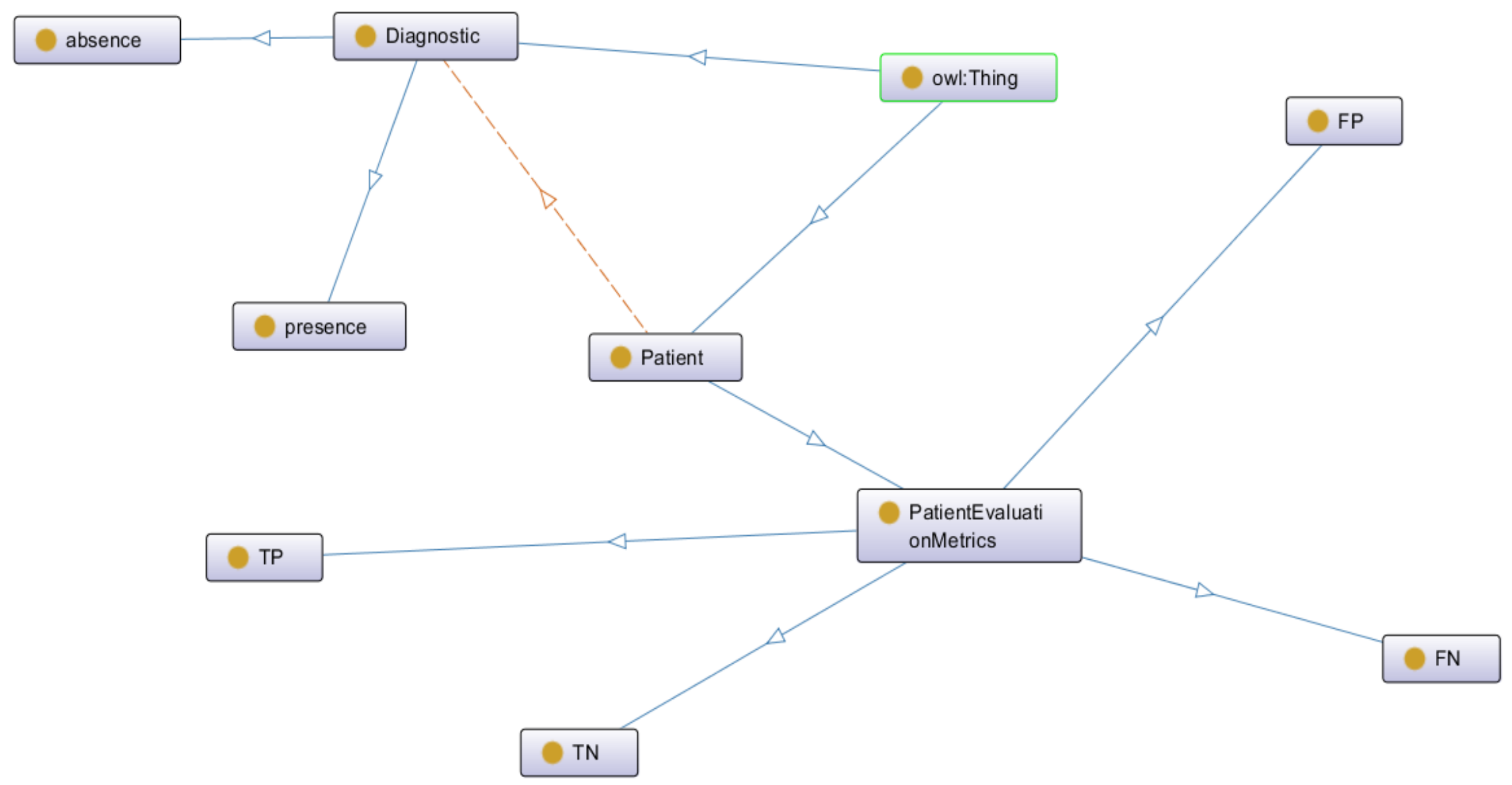

Figure 2. The ontology graphs

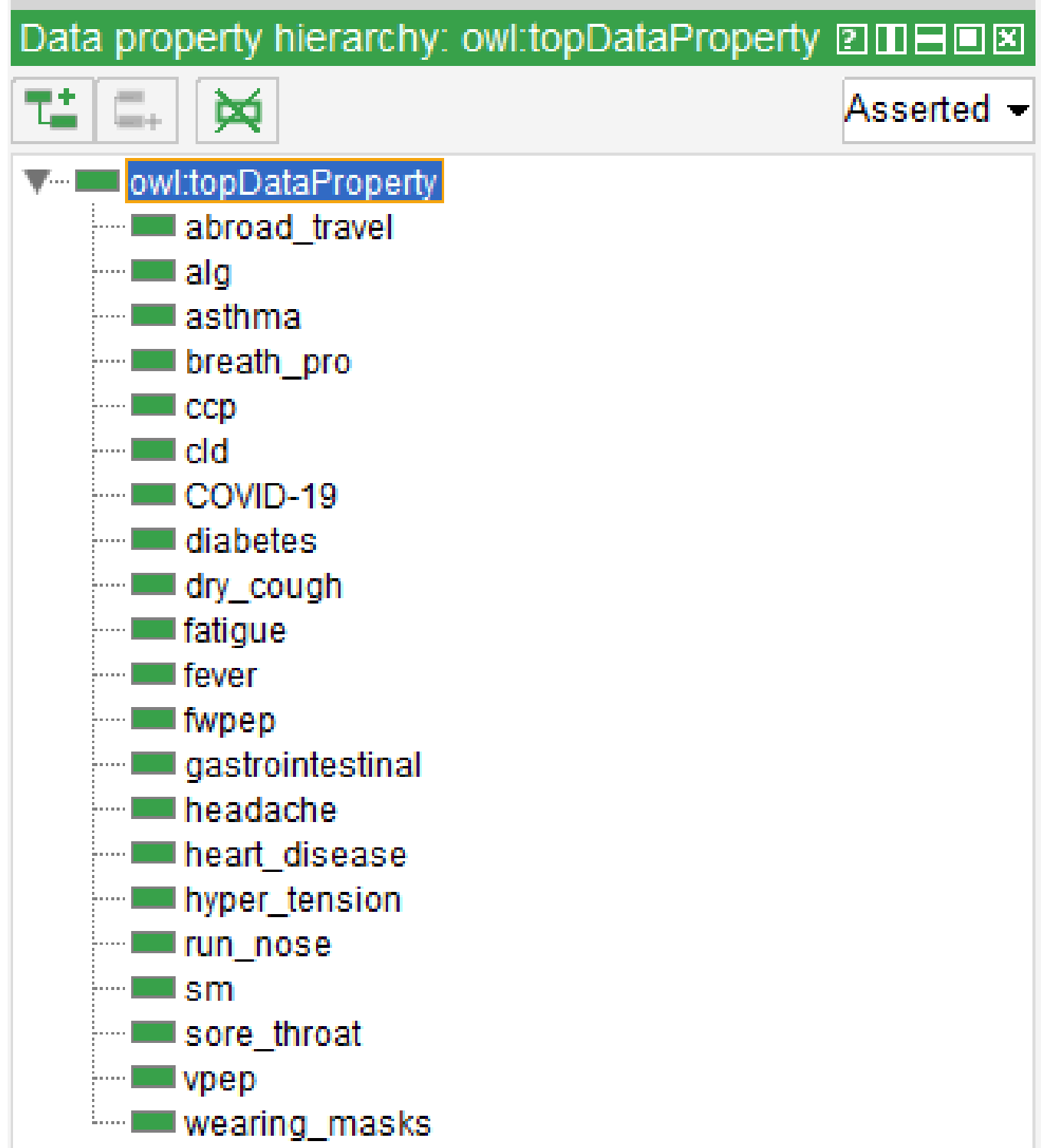

Figure 3. Data properties

To execute SWRL rules and infer new ontology axioms we utilized another plugin from Protégé software named Pellet [26], which includes capabilities for checking ontology coherence, deals with SWRL rules, computing the classification hierarchy, deals with OWL, explaining inferences, and answering SPARQL queries. It implements the ontology and SWRL rules to initiate the inference and then determines if the presence or absence of coronavirus disease. The ontology classifier's results are reported in the next section.

### 2.4. Evaluation

ROC area, F-measure, root mean squared error (RMSE), recall, accuracy, root relative squared error, precision, kappa statistic, and other performance measures are employed to assess ML algorithms. We employed two test modes (split-test and K-fold cross-validation) using several metrics including recall, F-Measure, Accuracy, and Precision to analyze our experimental results, which are presented below and in Figure 4. Furthermore, the same criteria are utilized to assess the validity of this comparison research including ML classifiers and the ontological model.

Other metrics, such as mean absolute error (MAE), MSE, and RMSE, are available but are most commonly employed in regression issues. As a result, owing to classification issues imposed by the dataset and techniques employed.

$$\text{ACC} = \frac{\text{TP+TN}}{\text{TP+TN+FP+FN}} \tag{1}$$

$$\text{PREC} = \frac{\text{TP}}{\text{TP+FP}} \tag{2}$$

$$\text{REC} = \frac{\text{TP}}{\text{TP+FN}} \tag{3}$$

$$\text{F-Measure} = 2 * \frac{\text{PREC}*\text{REC}}{\text{PREC+REC}} \tag{4}$$

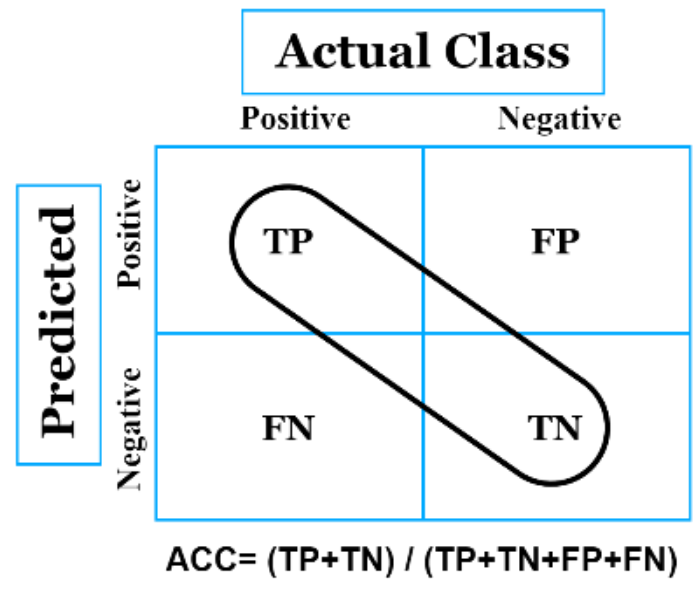

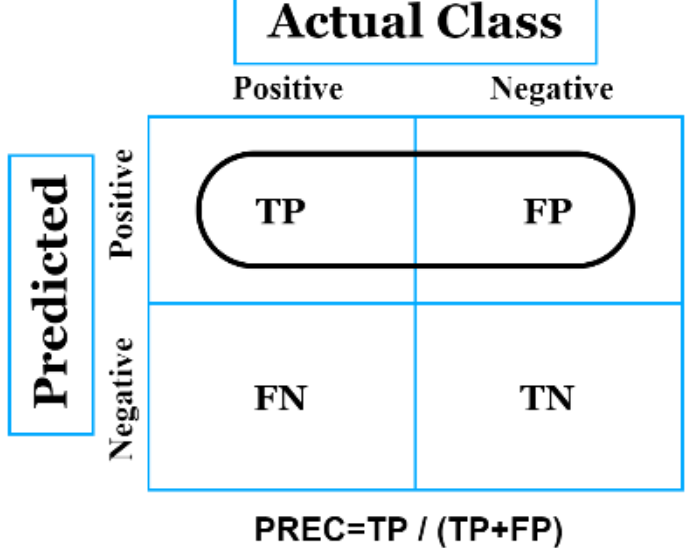

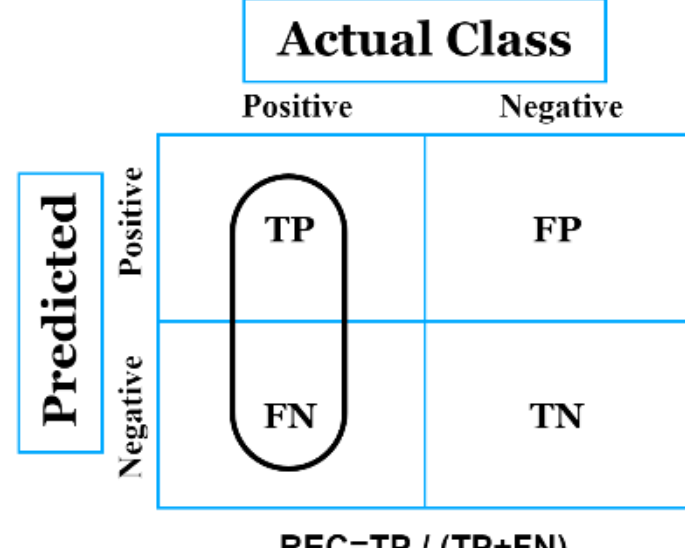

Figure 4. Performance metrics

## 3. RESULTS AND DISCUSSION

In this section, the results of the evaluation of the various classifiers that were used in this study are presented. The statistics and outcomes of the ontological model are also shown in Tables 2, 3, and Figure 5 illustrate the performance metrics of the ontology model, Figure 5(a) represents 10-fold cross-validation, and Figure 5(b) represents 70% split mode.

The results of this study provide a visual representation of the various metrics that are used in this research, such as precision, F-measure, Recall, and Accuracy, as shown in Figures 6 to 9. Table 4 also shows the results of the various classifiers that were used in this research.

Table 2. 10-fold cross-validation for ontological model

| Confusion matrix | | Actual class | |
|---|---|---|---|
| | | positive | negative |
| Predicted class | positive | TP: 389 | FP: 10 |
| | negative | FN: 2 | TN: 65 |

Table 3. 70% split mode for ontological model

| Confusion matrix | | Actual class | |
|---|---|---|---|
| | | positive | negative |
| Predicted class | positive | TP: 125 | FP: 1 |
| | negative | FN: 3 | TN: 11 |

Individuals by type (inferred):

absence (186)
FN (2)
FP (10)
presence (280)
TN (65)
TP (389)

(a)

Individuals by type (inferred):

absence (196)
FN (3)
FP (1)
presence (270)
TN (11)
TP (125)

(b)

Figure 5. Results of inferred concepts, (a) 10-fold cross-validation and (b) 70% split mode validation

### 3.1. Accuracy

The ontological model achieved the maximum value of 97.4% and SVM with rate of 96.8%, and 94.6% for both DT and Naïve bayes in terms of 10-fold cross-validation, according to the Figure 6 and Table 4. Almost the same results using split test mode, we obtained 97.1%, 96.4% for SVM, and 95.7% for both LR and DT.

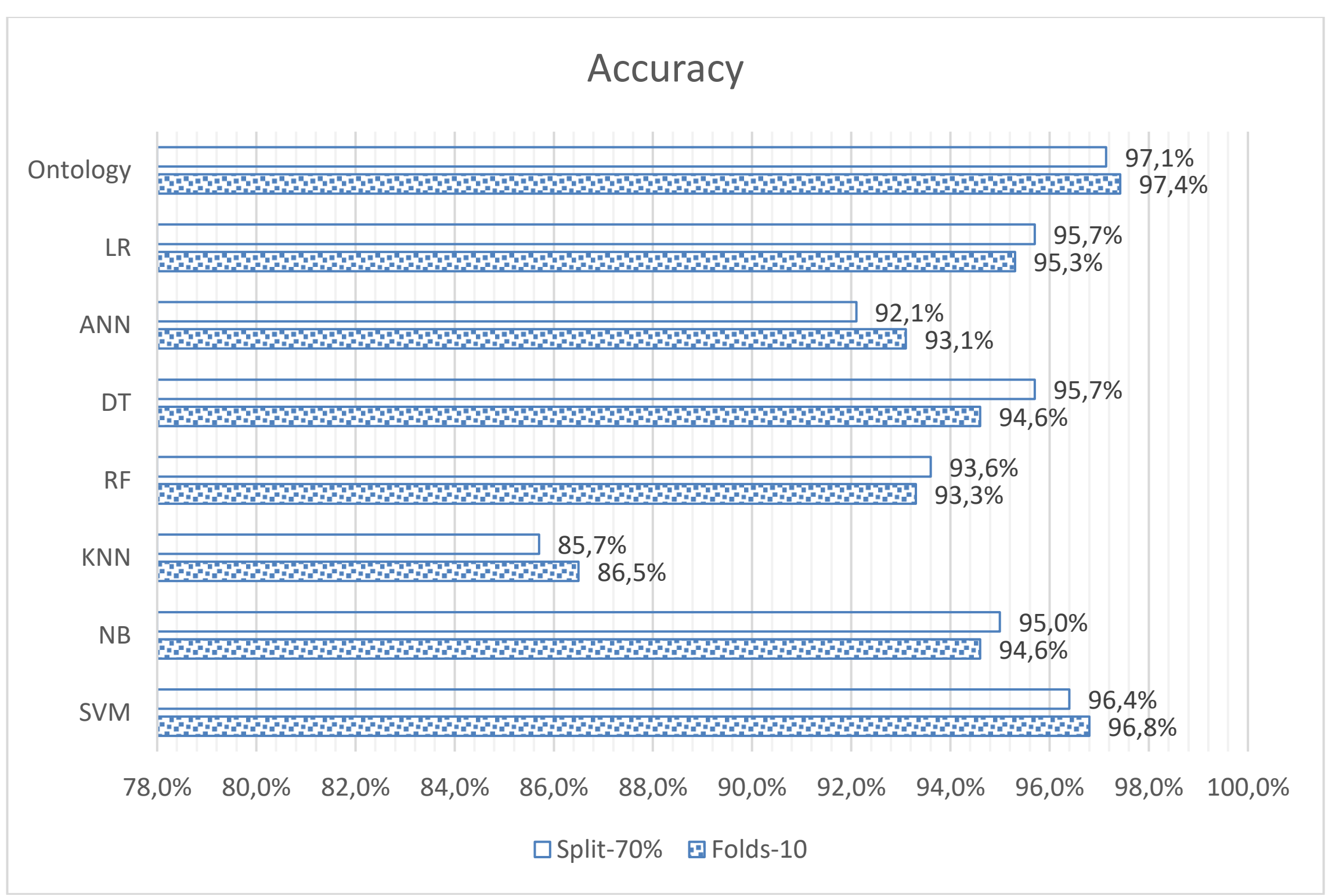

Figure 6. Accuracy comparison findings

### 3.2. Precision

The ontology classifier has the highest Precision of 99.2% in terms of split test mode, followed by SVM, LR, and DT. Concerning 10-fold cross-validation mode, the highest precision value of 97.5% goes for the ontology model. More details are shown in Table 4 and Figure 7.

### 3.3. Recall

According to Figure 8 and Table 4, the ontological model and SVM have the highest Recall values of 99.5% and 99.00% for Naïve Bayes, for 10-fold cross-validation mode. Concerning split test mode, the highest Recall value of 98.3% goes for Naïve bayes and SVM.

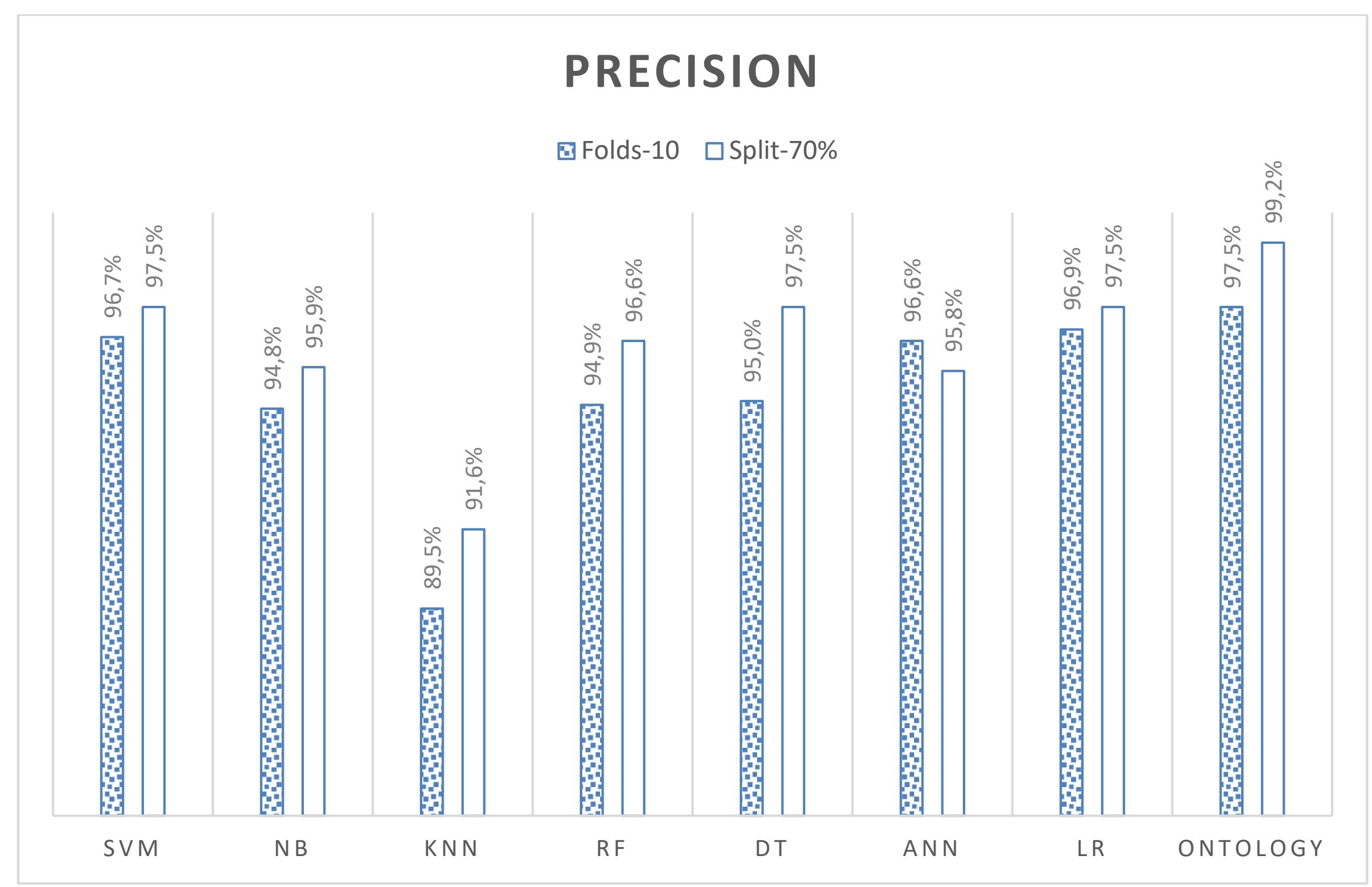

Figure 7. Comparison results of precision

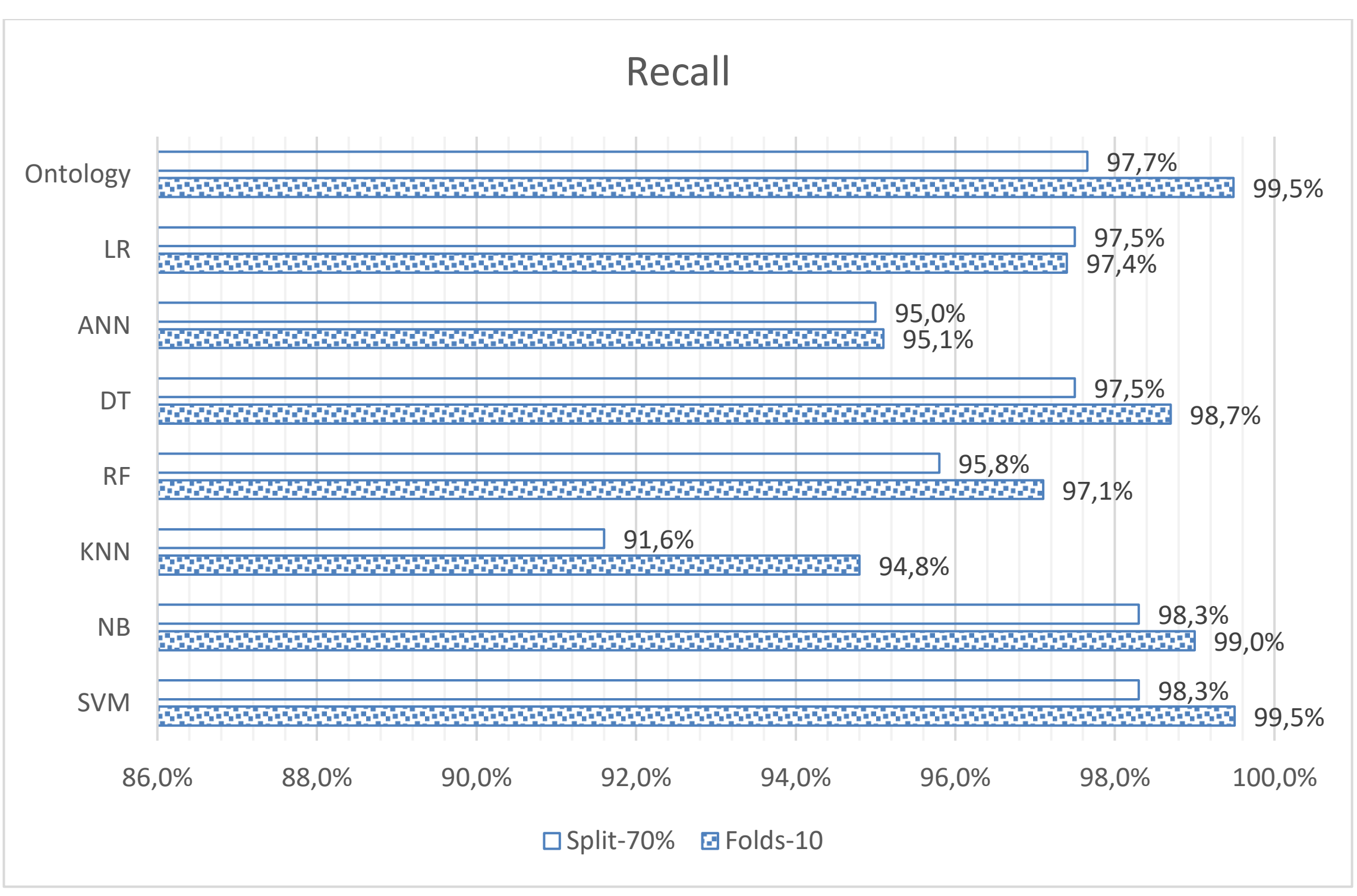

Figure 8. Recall comparison findings

### 3.4. F-measure

According to Figure 9 and Table 4, the ontology model had the greatest value of 98.5% in both test modes, followed by SVM in second position, and LR in third position.

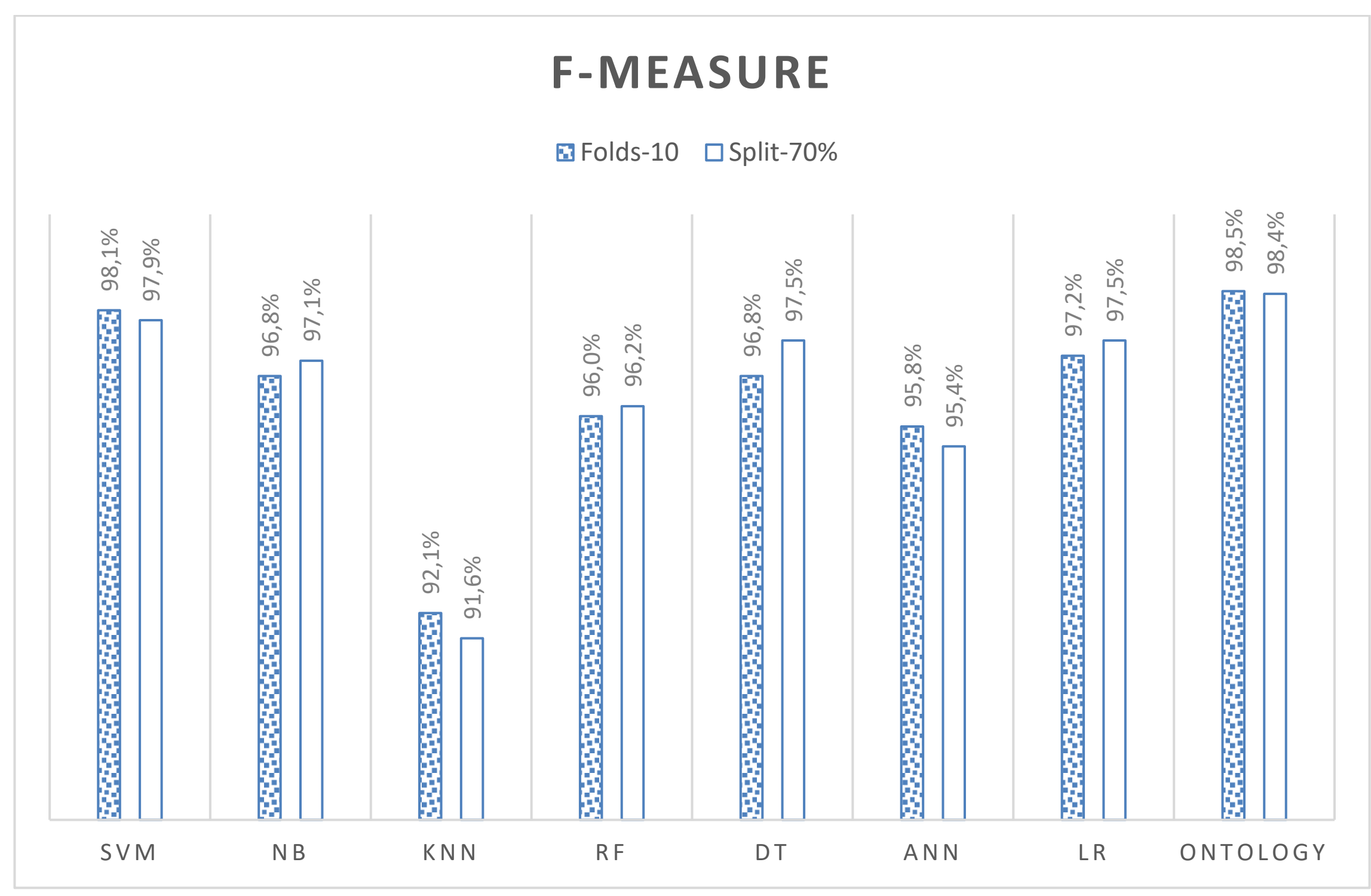

Figure 9. F-Measure comparison findings

Table 4. Ontological model and machine learning classifiers results

| | Accuracy | | Precision | | Recall | | F-Measure | |
|---|---|---|---|---|---|---|---|---|
| | Folds-10 | Split-70% | Folds-10 | Split-70% | Folds-10 | Split-70% | Folds-10 | Split-70% |
| SVM | 0.968 | 0.964 | 0.967 | 0.975 | 0.995 | 0.983 | 0.981 | 0.979 |
| NB | 0.946 | 0.95 | 0.948 | 0.959 | 0.99 | 0.983 | 0.968 | 0.971 |
| KNN | 0.865 | 0.857 | 0.895 | 0.916 | 0.948 | 0.916 | 0.921 | 0.916 |
| RF | 0.933 | 0.936 | 0.949 | 0.966 | 0.971 | 0.958 | 0.96 | 0.962 |
| DT | 0.946 | 0.957 | 0.95 | 0.975 | 0.987 | 0.975 | 0.968 | 0.975 |
| ANN | 0.931 | 0.921 | 0.966 | 0.958 | 0.951 | 0.95 | 0.958 | 0.954 |
| LR | 0.953 | 0.957 | 0.969 | 0.975 | 0.974 | 0.975 | 0.972 | 0.975 |
| Ontology | 0.974 | 0.971 | 0.975 | 0.992 | 0.995 | 0.977 | 0.985 | 0.984 |

The experimental findings reveal that the ontology model has the highest accuracy of 97.4%, followed by the SVM at 96.8%, LR at 95.3% and both DT and NB at 94.6%. In terms of the data stated above, we see no significant difference between 70%-Split and 10-Folds test mode. We conclude that, the ontological model can aid by extending the scope machine learning model. They can comprise any data kind or variation, and each diver data can be assigned to a certain job. Combining the ontological model with machine learning may provide well outcomes. The ontological model achieves results that are comparable to machine learning classifiers. Humans may interpret the findings, and the rules can be modified or added as needed. Furthermore, it supports unstructured, semi-structured, and structured data formats, allowing for more seamless data integration. It can comprise all aspects of the data modelling process, starting with schemas at the most basic level. As a result, they can handle the massive amounts of data utilized as input for machine learning training or output as outcomes. Furthermore, ontology matches any organization's aim, which might be mathematical, logical, or semantic-based. To the best of our knowledge, this is the first comparative study of the ontological model and ML in which we have integrated the ontology with ML, especially in the area of covid-19 prediction. As a result, no significant comparison can be done.

## 4. CONCLUSION

ML methods are widely employed in all scientific disciplines and have revolutionized industries all over the world. The use of machine learning techniques and algorithms in healthcare has recently advanced significantly. These approaches have shown success and may be valuable in the treatment of enduring diseases such as coronavirus. Furthermore, the Semantic Web has proven its usefulness and effectiveness in a multitude of areas, including health. As a Semantic Web component, ontology has the capability to treat concepts and relationships in the same way that humans view connected concepts.

In this research, we provided seven machine learning algorithms and an ontology model, as well as a comparison of their performance. Furthermore, two test modes are employed: 10-fold crossvalidation and percentage split, and several performance measures such as accuracy, F-measure, precision, and recall are employed to assess the outcomes. The findings show that the ontological model has the uppermost accuracy even when no feature selection is used. This brings us to a new search area, to which we advise and urge academics to participate and produce new insights in the same context, in order to provide additional outcomes and analysis, in order to make a forecast, recommendation, or decision, and so on. In future work, we want to improve this comparison analysis by adopting new ways to incorporate ML rules with the ontological model method, as well as regression machine learning algorithms.

## BIOGRAPHIES OF AUTHORS

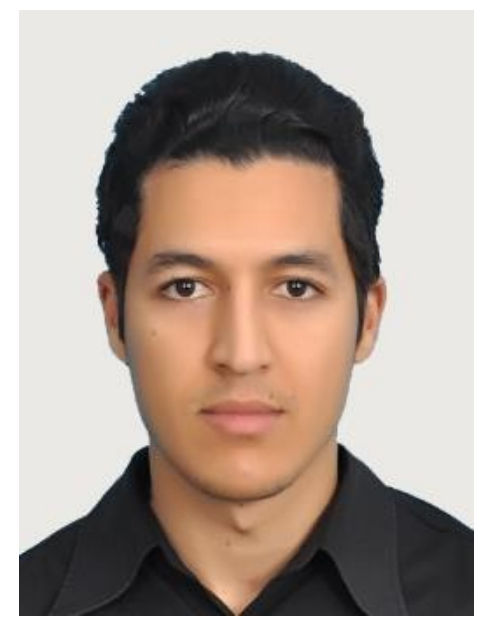

**Hakim El Massari** received his master degree from Normal Superior School of Abdelmalek Essaadi University, Tétouan, Morocco, in 2014. Currently, he is preparing his Ph.D. in computer science at the National School of Applied Sciences, Sultan Moulay Slimane University, Khouribga, Morocco. His research areas include machine learning, deep learning, big data, semantic web, and ontology. He can be contacted at email: h.elmassari@usms.ma.

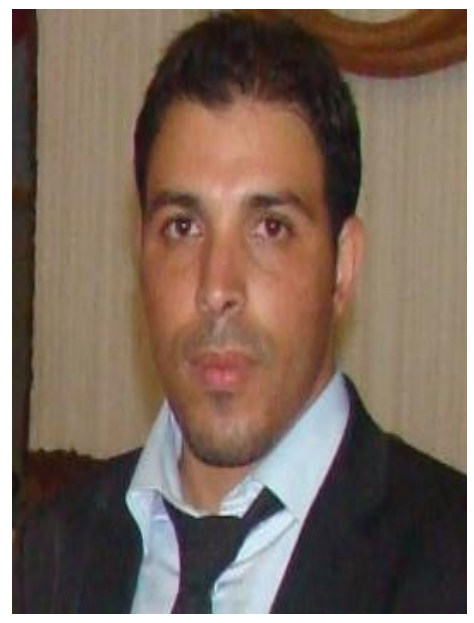

**Noreddine Gherabi** is a professor of computer science with industrial and academic experience. He holds a doctorate degree in computer science. In 2013, he worked as a professor of computer science at Mohamed Ben Abdellah University and since 2015 has worked as a research professor at Sultan Moulay Slimane University, Morocco. Member of the International Association of Engineers (IAENG). Professor Gherabi has several contributions in information systems namely: big data, semantic web, pattern recognition, intelligent systems. He has papers (book chapters, international journals, and conferences/workshops), and edited books. He has served on executive and technical program committees and as a reviewer of numerous international conferences and journals, he convened and chaired more than 30 conferences and workshops. He is member of the editorial board of several other renowned international journals:

• Co-editor in chief (Editorial Board) in the journal "The International Journal of sports science and engineering for children" (IJSSEC).

• Associate Editor in the journal « International Journal of Engineering Research and Sports Science ».

• Reviewer in several journals/Conferences

• Excellence Award, the best innovation in science and technology 2009

His research areas include machine learning, deep learning, big data, semantic web, and ontology. He can be contacted at email: n.gherabi@usms.ma.

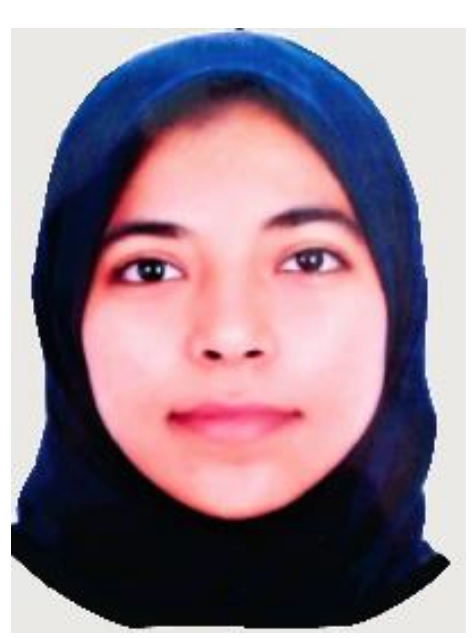

**Sajida Mhammedi** received her Ms Degree in Computer Engineering from Faculty of Science and Technologie, Beni Mellal Morocco, she worked as a visiting researcher at the Sultane Moulay Slimane University, her research interests include machine learning, semantic web, recommendation systems, ontology, and big data. She can be contacted at email: sajida.mhammedi@usms.ac.ma.

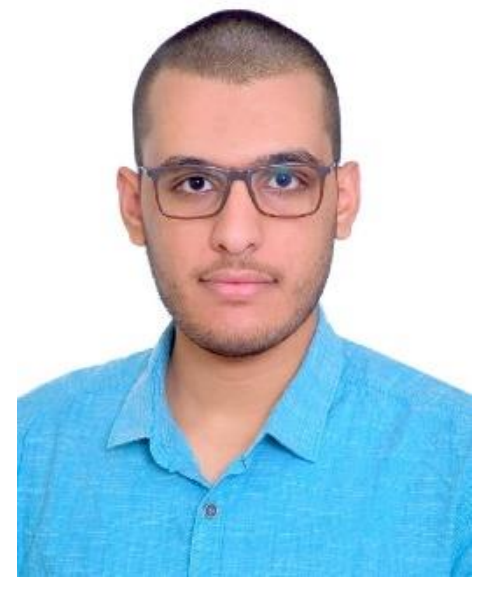

**Hamza Ghandi** is pursuing her Ph.D. in Computer Engineering from the National School of Applied Sciences of Khouribga, her area of interest is machine learning, intelligent systems, deep learning, and big data. He can be contacted at email: hamza.ghandi@usms.ac.ma.

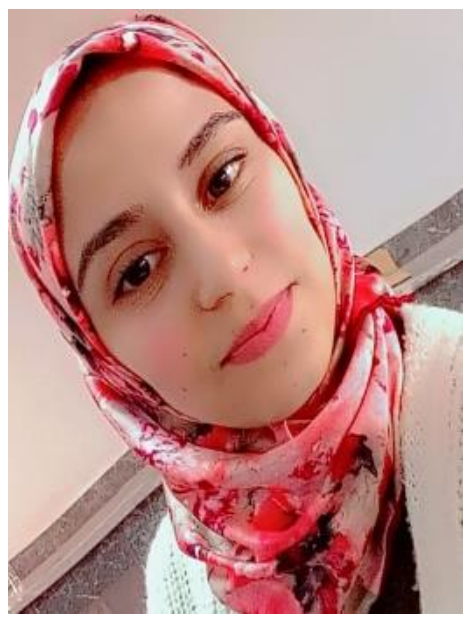

**Fatima Qanouni** is pursuing her Ph.D. in Computer Engineering from the National School of Applied Sciences of Khouribga, her area of interest is machine learning, intelligent systems, deep learning, and big data. She can be contacted at email: Fatima.qanouni@usms.ac.ma.

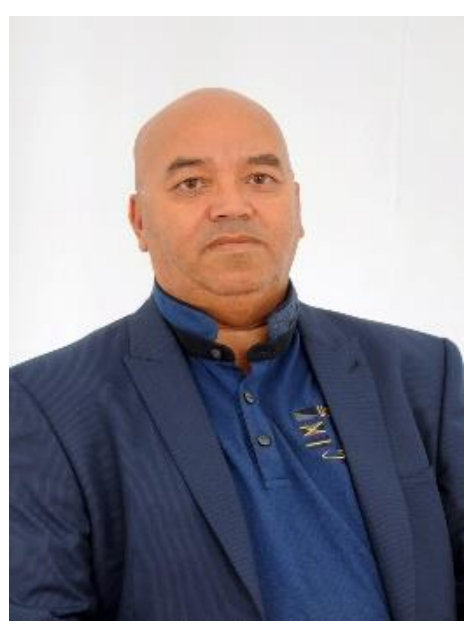

**Mohamed Bahaj** obtained his Ph.D. in Mathematics and Computer Science from University Hassan 1st, Morocco. He is co-chairs of International Conferences on International Conference on Intelligent Information and Network Technology. He is a Full Professor in Department of Mathematics and Computer Sciences from the University Hassan 1st Faculty of Sciences & Technology Settat Morocco. He has published over 120 peer-reviewed papers. His research interests are Intelligent Systems, Ontologies Engineering, Partial and differential Equations, Numerical Analysis and Scientific Computing He has reviewed several papers in various journals. He has supervised several PhD thesis in Computers Sciences & in Mathematics. He is a Guest Editorial in Journal of Emerging Technologies in Web Intelligence. He also attended a series of workshops, seminars and discussion forums for Academic Development on teaching and research. He can be contacted at email: mohamedbahaj@gmail.com.